\documentclass[aps,twocolumn,showpacs,superscriptaddress,amsmath,amssymb,amsfonts,floatfix,longbibliography]{revtex4-1}

\usepackage{graphicx}
\usepackage{dcolumn}
\usepackage{bm}
\usepackage{amssymb}
\usepackage{microtype}
\usepackage{xfrac}
\usepackage{gensymb}
\usepackage[most]{tcolorbox}
\usepackage{xcolor}
\usepackage{multirow}
\usepackage{enumitem}
\usepackage{natbib,hyperref}
\usepackage{graphicx}
\hypersetup{
	colorlinks=true, 
	linkcolor=blue,  
}

\begin{document}

\setlength{\parskip}{0pt}

\title{Tutorial: A Beginner's Guide to Interpreting Magnetic \\ Susceptibility Data with the Curie-Weiss Law} 

\author{Sam Mugiraneza}
\affiliation{Quantum Matter Institute, University of British Columbia, Vancouver, BC V6T 1Z4, Canada}
\affiliation{Department of Chemistry, University of British Columbia, Vancouver, BC V6T 1Z1, Canada}

\author{Alannah M. Hallas}
\email[Email: ]{alannah.hallas@ubc.ca}
\affiliation{Quantum Matter Institute, University of British Columbia, Vancouver, BC V6T 1Z4, Canada}
\affiliation{Department of Physics \& Astronomy, University of British Columbia, Vancouver, BC V6T 1Z1, Canada}

\begin{abstract}

    Magnetic susceptibility measurements are often the first characterization tool that researchers turn to when beginning to assess the magnetic nature of a newly discovered material. Breakthroughs in instrumentation have made the collection of high quality magnetic susceptibility data more accessible than ever before. However, the analysis of susceptibility data remains a common challenge for newcomers to the field of magnetism. While a comprehensive treatment of the theoretical aspects of magnetism are found in numerous excellent textbooks, there is a gap at the point of practical application. We were inspired by this obstacle to put together this guide to the analysis and interpretation of magnetic susceptibility data, with an emphasis on materials that exhibit Curie-Weiss paramagnetism. 
    
\end{abstract}

\maketitle

\section*{Introduction}

Magnetic materials are of tremendous importance in both fundamental science and in applications-driven research. The experimental toolkit for characterizing magnetic materials continues to grow year after year, with advanced new experiments providing remarkable insights~\cite{tetienne2014nanoscale,kasahara2018majorana,thiel2019probing,yavacs2019direct}. However, for newly synthesized materials, there is one indispensable characterization technique that is as old as the field of magnetism itself: magnetic susceptibility, $\chi$,
\begin{equation} \label{Eq2:susceptibility}
M = \chi H,
\end{equation} 
which is the quantity that relates a material's magnetization, $M$, to the strength of an applied magnetic field, $H$. In this tutorial we will proceed with the conventional usage, which assumes that these two quantities are linearly related, which is typically most valid at high temperatures and low fields. In very general terms, magnetic susceptibility measurements tell you how your material responds to an applied magnetic field, which can be used to unveil the magnetic identity of your material. Susceptibility measurements can be performed in a direct current (DC) field, giving insights on the static magnetic properties, or an alternating current (AC) field in order to probe the dynamic properties.

For more than a hundred years, the preferred method for measuring magnetic susceptibility was based on determining the apparent weight of a sample in a magnetic field, as in the Gouy and Faraday balance techniques~\cite{mckeehan1929measurement, kapitza1931method,morris1968faraday}. However, in the modern era, more precise measurements have been unlocked through the advent of the SQUID (Superconducting Quantum Interference Device, see Box 1) magnetometer and its subsequent commercialization~\cite{jaklevic1964quantum,clarke2006squid,fagaly2006superconducting}. Procuring high quality temperature dependent magnetic susceptibility measurements down to liquid helium temperatures, or colder, has therefore never been easier. At the data collection stage, your instrument's user manual is an important resource, that can guide you on the optimal conditions under which to perform your measurement. There is, however, no equivalent, compact resource for the next step: analyzing and interpreting your data set. Instead, one must often resort to haplessly skimming through the literature hoping to stumble upon data that resembles their own. Our objective here is to fill that gap by cataloging the range of behaviors that can arise in magnetic susceptibility measurements, particularly in the temperature range above any ordering transition.

In this tutorial, we provide a guide to the interpretation of magnetic susceptibility data with a special emphasis on the Curie-Weiss law, a simple but powerful equation. Our hope is that this tutorial can supplement many excellent and comprehensive textbooks on the magnetism of solids~\cite{blundell2003magnetism,mohn2006magnetism,spaldin2010magnetic,coey2010magnetism}, which lay the foundation for the topics discussed here. In Box 1 we provide a brief glossary for some key terms that will be used throughout. Our discussion here explicitly focuses on susceptibility measured with a DC field, as this is the most routine type of measurement for the characterization of new materials. All data presented herein has been measured with an applied DC field. However, much of the theoretical underpinnings discussed in the sections that follow apply equally to DC and AC susceptibility measurements. Indeed, in the limit where there are no low energy excitations, measurements in an AC and DC field are equivalent. When this limit does not hold, AC susceptibility can provide additional insights~\cite{balanda2013ac,topping2018ac}.

\vspace{-2mm}

\section*{Magnetism units}
\label{units}

Unit systems are a common stumbling block for newcomers to the field of magnetism. While the cgs (centimeter-gram-second) system is more prevalent, some sources prefer SI (international system) units. Converting between these two unit systems in the context of magnetism is a particularly error prone activity with several subtleties. It is important to follow the conventions of your field and also to familiarize yourself with conversions for the most frequently used units. For example, molar susceptibility data are commonly  represented in emu\,mol$^{-1}$ (cgs), which can be converted into m$^3$\,mol$^{-1}$ (SI), by multiplying by a factor of $4\pi \cdot 10^{-6}$. Note that, in cgs units, emu\,Oe$^{-1}$ is equivalent to emu and thus Oe is often omitted from susceptibility units in the axes labels of published figures. The appendices in several books have useful tables for SI-cgs conversion~\cite{blundell2003magnetism,coey2010magnetism,cullity2011introduction} and Bennett \emph{et al.} provides an interesting historical perspective~\cite{bennett1978comments}. Here we will primarily proceed with cgs units, as these are default units of most commercial SQUID magnetometers and also the unit system that one more commonly encounters in the physics literature, but key equations are provided in both cgs and SI unit systems.

\vspace{4mm}

\begin{tcolorbox}[breakable]
\subsection*{Box 1: Glossary of important terms for magnetic susceptibility measurements}
\begin{itemize}[leftmargin=*]
    
    \item \textbf{SQUID magnetometer}: a high sensitivity instrument that uses a SQUID (an acronym for Superconducting QUantum Interference Device) to enable the direct measurement of a material's magnetic susceptibility. A SQUID is a superconducting loop with two parallel Josephson junctions that can detect incredibly small changes in magnetic flux. The two other components that make up the SQUID magnetometer are a superconducting magnet and a cryostat, which allow field and temperature dependent susceptibility measurements to be performed.

    \item \textbf{Susceptibility} and \textbf{magnetization}: colloquially, these terms are often used interchangeably to describe the magnetic field induced in a material through the application of an external magnetic field. Generally, the term susceptibility is preferred for measurements performed with varying temperature and fixed applied field while magnetization is more often used to refer to measurements performed at fixed temperature with varying applied field.

    \item \textbf{DC} and \textbf{AC susceptibility}: measurements of magnetic susceptibility can be collected with a DC (direct current) or AC (alternating current) applied field. In a DC measurement, the field is held constant and the measurement is obtained under equilibrium conditions. Conversely, in an AC susceptibility measurement, the field oscillates inducing a time-dependent magnetization that provides insights on the material's dynamics. In an AC measurement, the response can also depend upon the frequency at which the field is alternated.

    \item \textbf{Paramagnetic} and \textbf{diamagnetic}: these terms generally refer to the sign of the susceptibility as related to the direction of the external magnetic field, with paramagnetic referring to a positive susceptibility (net moment in the same direction as the applied field) and diamagnetic referring to a negative susceptibility (net moment opposing the magnetic field).

    \item \textbf{Field cooled (FC)} and \textbf{zero field-cooled (ZFC)}:  the initial state of a material in a susceptibility measurement can depend on whether the sample was measured under FC or ZFC conditions, which describe whether the sample was cooled down in the presence or absence of an applied magnetic field, respectively. Divergence between measurements performed in FC and ZFC conditions, often termed splitting, indicates hysteresis (dependence on the magnetic field history), which can be indicative of various ordered and frozen magnetic states. 

    \item \textbf{Isotropic} and \textbf{Anisotropic}: these terms are used to distinguish material properties that depend on measurement direction from those that do not. In the case of magnetic susceptibility, this  means the crystallographic direction along which the magnetic field is applied. The most common origin of anisotropy is crystal field effects that can constrain the moments to lie along specific axes or within a specific plane. Detecting anisotropy generally requires single crystal samples since in a polycrystalline sample all directions are averaged over. 
    
    \item \textbf{Exchange interactions}: the quantum mechanical mechanism through which neighboring magnetic ions interact with one another, which can ultimately result in long-range magnetic order. This exchange can occur directly, when the orbitals of the magnetic ions themselves overlap, or indirectly, through an adjoining ligand such as oxygen via a superexchange mechanism, or by the conduction electrons, which is known as the RKKY (Ruderman–Kittel–Kasuya–Yosida) interaction.

\end{itemize}

\end{tcolorbox}

\section*{Contributions to magnetic susceptibility}
\label{contributions}

In this section we briefly introduce the various diamagnetic and paramagnetic contributions to magnetic susceptibility that can arise depending upon the magnetic and electronic character of the material in question. We begin with orbital (or core) diamagnetism, which occurs in all materials, before moving on to the local moment contributions to magnetism: Curie-Weiss paramagnetism and van Vleck paramagnetism. We then briefly outline the contributions that arise in metallic materials: Pauli paramagnetism, Landau diamagnetism, and the temperature dependent paramagnetic response observed in itinerant moment systems. Finally, we describe the characteristic forms of various types of ordering or freezing transitions, which can be detected with magnetic susceptibility measurements. It is worth emphasizing that these contributions are not mutually exclusive; for example, a magnetic rare earth metal will have Curie-Weiss paramagnetic, Pauli paramagnetic, and core diamagnetic terms in its susceptibility.

\subsection*{Orbital diamagnetism}

Orbital diamagnetism, which describes the tendency of electrons to repel a magnetic field giving rise to a negative susceptibility, is a property of all materials. In a classical framework, the origin of diamagnetism is often discussed in terms of Lenz's law, which states that an applied field on the orbital motion of an electron induces a current that opposes the applied field. However, a rigorous derivation of the diamagnetic response of electrons requires a quantum mechanical description based on first order perturbation theory~\cite{morrish2001physical,blundell2003magnetism}. The diamagnetic susceptibility, $\chi_{\text{D}}$, in units of emu\,mol$^{-1}$ [cgs] or m$^3$\,mol$^{-1}$ [SI] is given by
\begin{equation}
  \chi_{\text{D}} = -\frac{n e^2}{6m_{\text{e}}c^2}\langle r^2 \rangle \; \text{[cgs]} = -\frac{n \mu_0 e^2}{6m_{\text{e}}}\langle r^2 \rangle \; \text{[SI]},
\end{equation}\label{Eq5:orbital diamagnetism}
where $n$ is the number of electrons per mole, $\mu_0$ is the vacuum permeability, $e$ is the elementary charge, $\langle r^2 \rangle$ is the average square radius of the electron orbit, $m_{\text{e}}$ is the electron mass, $c$ is the speed of light in vacuum, and the negative sign reflects the repulsive tendency. While diamagnetism  occurs in all materials, it is a weak effect ($-10^{-6}$ to $-10^{-5}$ emu\,mol$^{-1}$) that is typically overshadowed by any other contribution to the magnetic susceptibility. Therefore, the label diamagnet is reserved for materials where diamagnetism is the only contribution to the susceptibility, typically insulators with no unpaired electrons. 

In many cases, the diamagnetic contribution to the susceptibility is so small that it can simply be ignored. However, in certain situations, it may be desirable to correct the measured data by subtracting off the diamagnetic term. This is a relatively straightforward task because the diamagnetic contribution is temperature independent and can be accurately approximated using tabulated values for the constituent ions~\cite{bain2008diamagnetic}. The total diamagnetic susceptibility is then simply the sum over all atoms. While this approach provides a reasonable estimate, it's worth emphasizing that these tables assume purely covalent or purely ionic environments while real materials often exist between these two extremes.

\begin{figure} 
  \centering
  \includegraphics[width= 8.5cm]{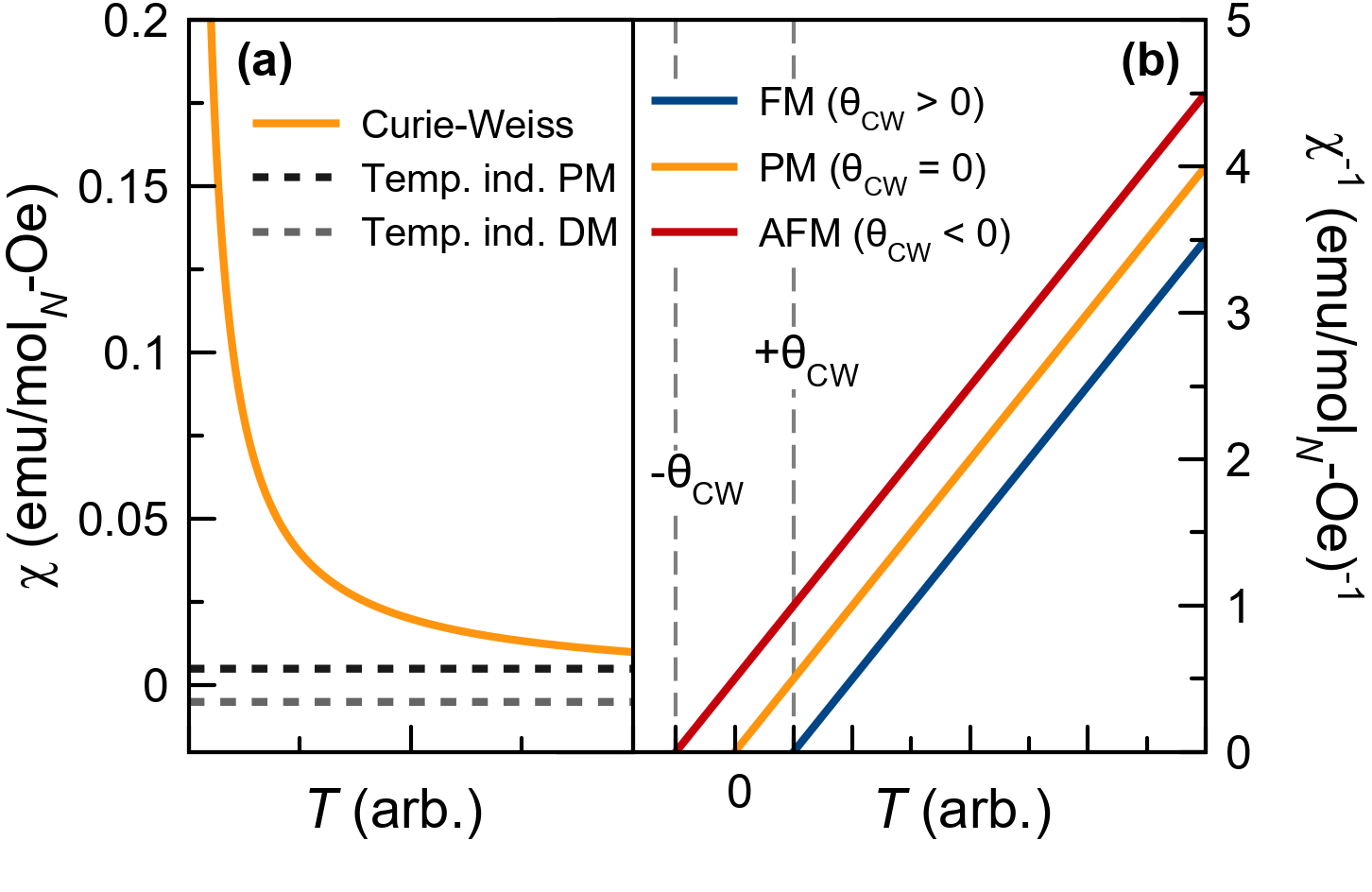}
  \caption{Appearance of Curie-Weiss behavior in direct and inverse susceptibility. \textbf{(a)} Curie Weiss susceptibility due to paramagnetic local moments goes as $\sfrac{1}{T}$ and is typically at least an order of magnitude larger than than other temperature independent contributions to susceptibility such as Pauli or van Vleck paramagnetism (PM), or orbital and Landau diamagnetism (DM). \textbf{(b)} The inverse susceptibility of a material that follows the Curie-Weiss law  will be linear in temperature and the $x-$intercept gives the Curie-Weiss temperature ($\theta_{\text{CW}}$) with positive values indicating net ferromagnetic (FM) interactions, negative values indicating antiferromagnetic (AFM) interactions, and values close to zero indicating negligible interactions.}
  \label{CurieWeissSchem}
\end{figure}

\subsection*{Curie-Weiss paramagnetism}
\label{CW_paramagnetism}

Unlike the diamagnetic response which occurs in all materials, paramagnetism occurs exclusively in ``magnetic'' materials – materials with unpaired electrons. At some temperature commensurate with the strength of the magnetic correlations, these magnetic materials may undergo a symmetry breaking transition to a magnetically ordered state. However, at relatively higher temperatures, the material exists in magnetically disordered gas-like state known as a paramagnet, where thermal fluctuations are stronger than the interactions between magnetic ions. There is much that can be learned from studying the susceptibility of a paramagnet. In particular, a mean field treatment of this state gives the equation at the heart of this paper, the Curie-Weiss law.

The Curie-Weiss law, which is derived as an extension of Curie’s law by incorporating the concept of Weiss’s molecular field is
\begin{equation}
\chi = \frac{C}{T-\theta_{\text{CW}}} \; \text{[cgs and SI]},
\end{equation}
where $C$ is known as the Curie constant (with units emu\,K\,mol$^{-1}$ in cgs or m$^3$\,K\,mol$^{-1}$ in SI) and $\theta_{\text{CW}}$ (with units K) is often referred to as the Curie-Weiss temperature. This equation captures the tendency for the moments in a paramagnet to align with the external field. This results in a susceptibility that increases with decreasing temperature, as thermal fluctuations become less potent (see Fig.~\ref{CurieWeissSchem}(a)). The Curie constant $C$ is directly related to the number of unpaired electrons and, once determined, can be used to calculate the effective magnetic moment per ion in units of Bohr magnetons, $\mu_{\text{B}}$, 
\begin{equation} \label{effective}
\mu_{\text{eff}} = \sqrt{8C} \; \mu_{\text{B}} \; \text{[cgs]} \simeq 800\sqrt{C} \; \mu_{\text{B}} \; \text{[SI]}.
\end{equation}
This effective moment can be directly compared to the calculated value for the ion in question, given by
\begin{equation} \label{mucalc}
\mu_{\text{cal}} = g_J \sqrt{J(J+1)} \; \mu_{\text{B}},
\end{equation}
which depends only on its total angular momentum $J$ and its g-tensor $g_J$ (for more on determining $J$ refer to the sections on Hund’s rules in your preferred magnetism textbook~\cite{coey2010magnetism,morrish2001physical, blundell2003magnetism}). It's worth emphasizing that the derivation of the Curie-Weiss law presupposes the existence of a well-defined angular momentum ground state. 

The magnitude of the Curie-Weiss temperature $\theta_{\text{CW}}$ is related to the strength of the molecular field, which can be taken as an approximate indicator of the strength of the magnetic correlations between ions. Positive values of $\theta_{\text{CW}}$ occur when the molecular field aligns with the external field, indicating ferromagnetic interactions, whereas negative values of $\theta_{\text{CW}}$ imply antiferromagnetic interactions (Fig.~\ref{CurieWeissSchem}(b)). As the temperature grows close to $|\theta_{\text{CW}}|$, the mean field treatment breaks down and deviations from the Curie-Weiss law are expected, which could include a transition to a magnetically ordered or frozen state (see Box 2). For ferromagnets, one often finds that $T_{\text{C}} \approx \theta_{\text{CW}}$ whereas larger deviations are typically observed for antiferromagnets, $T_{\text{N}} < |\theta_{\text{CW}}|$, due to an oversimplification of how the molecular field is defined. In some cases, the value is even further suppressed due to the effect of frustration, as will be discussed later in this tutorial. 

The best adherence to Curie-Weiss behavior is encountered in $4f$ rare earth compounds and insulating $3d$ transition metal compounds, which will be discussed in detail below. In the former case, this is because the $4f$ electron orbitals are highly spatially localized and thus, even in metallic materials, they are buried so deep below the conduction band that the magnetic moments are completely localized. In the case of $3d$ transition metals, Mott insulating states are often observed due to the dominant effect of Coulomb repulsion giving rise to well localized $3d$ magnetic moments. These are the two classes of materials we will therefore focus our remarks on here. However, it's worth emphasizing that good realizations of Curie-Weiss behavior can also be found elsewhere on the periodic table, including among $4d$ and $5d$ transition metal compounds. With these more spatially extended orbitals there is an increasing tendency towards metallicity, with the partially filled $d$ band constituting the conduction band, leading to the breakdown of the Curie-Weiss description.

\begin{figure*}
  \centering
  \includegraphics[width=18cm]{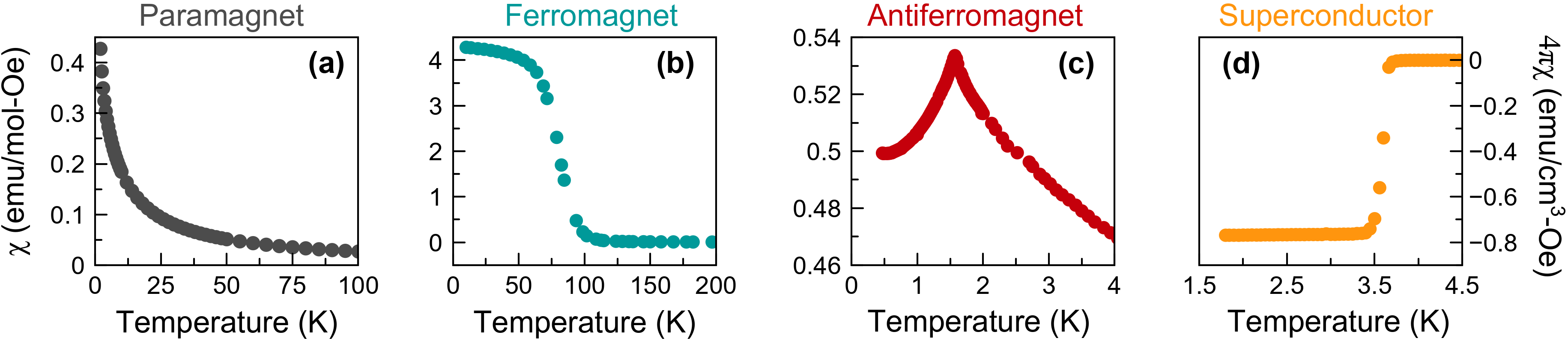}
  \caption{Magnetic susceptibility measurements are sensitive to the presence (or absence) of long-range ordering transitions, as shown in these reproduced data sets. \textbf{(a)} Materials with dilute magnetic moments may remain paramagnetic down to the lowest temperatures, such as this data set for dilute Mn$^{2+}$ moments in SrMn$_{\sfrac{1}{2}}$Te$_{\sfrac{3}{2}}$O$_6$~\cite{hallas2019sr}. \textbf{(b)} Ferromagnetic (and ferrimagnetic) transitions are marked by a sharp increase in the susceptibility as the moments align with the external field, as shown for Ni$_{0.68}$Rh$_{0.32}$ with $T_{\text{C}} = 96$~K~\cite{huang2020quantum}. \textbf{(c)} Antiferromagnetic transitions are marked by a sharp cusp in the susceptibility, as shown for Gd$_2$Pt$_2$O$_7$ with $T_{\text{N}} = 1.6$~K~\cite{hallas2016relief}. \textbf{(d)} Superconducting transitions are marked by an abrupt decrease in the susceptibility due to the perfect diamagnetic screening in the superconducting states, as shown for PbTaSe$_2$~\cite{wilson2017mu}.} 
  \label{Transitons}
\end{figure*}

\subsection*{van Vleck paramagnetism}

Our discussion of paramagnetism thus far has considered only the affect of an applied magnetic field on electrons in their angular momentum ground state. However, in all cases where there are partially filled electron orbitals, higher energy angular momentum states do exist. A correction of the quantum mechanical treatment used to obtain the expression for core diamagnetism with second order perturbation theory yields an additional positive term that depends on these excited states known as van Vleck paramagnetism. This temperature independent term is inversely proportional to the energy gap between the angular momentum states. In the vast majority of materials, this energy gap is so large that, at experimentally relevant temperatures, van Vleck paramagnetism can be entirely ignored. The most notable exceptions are compounds based on Eu$^{3+}$ and Sm$^{3+}$, which have energy gaps to the first excited state that are comparable to thermal energy at room temperature.

\subsection*{Pauli paramagnetism and Landau diamagnetism} 

The susceptibility of non-localized conduction electrons, which by definition is only relevant to metals, is called Pauli paramagnetism. In the absence of a magnetic field, the number of spin up and spin down electrons in the partially filled conduction band are equal. However, the application of a magnetic field breaks the degeneracy of the spin-up and spin down bands, resulting in a small imbalance of spins aligned with the external field, yielding a  positive susceptibility. The Pauli susceptibility, $\chi_{\text{P}}$, is well-approximated  by the notably temperature-independent expression
\begin{equation} \label{Eq7:Puali paramagnetism}
\chi_{\text{P}} = \mu_{\text{B}}^2 g(E_{\text{F}}) \; \text{[cgs]} = \mu_0 \mu_{\text{B}}^2 g(E_{\text{F}}) \; \text{[SI]},
\end{equation}
which depends only on the density of states at the Fermi energy, $g(E_{\text{F}})$. The Pauli susceptibility is typically weak, of order $10^{-4}$ to $10^{-5}$ emu\,mol$^{-1}$, because the applied field only affects the small fractions of electrons that are close to the Fermi level \cite{cullity2011introduction}.

The associated orbital contribution to the susceptibility of conduction electrons is called Landau diamagnetism, $\chi_{\text{L}}$. In a free electron model, the Landau diamagnetic response is precisely one third of the Pauli paramagnetic response, but with opposite sign. However, the ratio of these two terms is sensitive to details of the band structure and can vary significantly in cases where the effective mass of the conduction electrons, $m^*$, is significantly different from the free electron mass, $m_{\text{e}}$.

\subsection*{Temperature-dependent itinerant moment paramagnetism}

While the paramagnetic response of conduction electrons is typically small and temperature independent, there are a special subset of materials that buck this trend due to the presence of intense electron-electron correlations. These materials are known as itinerant magnets. In contrast to local moment magnetism, which arises in individual atoms with unpaired electrons, itinerant moment magnetism is an inherently collective behavior originating from the electron bands which cross the Fermi energy~\cite{moriya2012spin,kubler2017theory,santiago2017itinerant}. As is the case for a local moment system, itinerant magnets can undergo a magnetic ordering transition at a temperature characteristic of the strength of the interactions. It is, however, worth emphasizing that itinerant magnets are vastly outnumbered by local moment systems and that many real materials are found to exist in an intermediate regime with dual local-itinerant character.

While the physical origin of local and itinerant moments are distinct, a striking fact is that itinerant magnets also exhibit a Curie-Weiss like susceptibility: that is to say, $\chi$ is inversely proportional to temperature above the magnetic ordering transition. This effect originates from the temperature dependence of the mean square local amplitude of the spin fluctuations, as first described by Moriya~\cite{moriya2012spin}. Thus, while one can fit the paramagnetic susceptibility of an itinerant magnet to the standard Curie-Weiss equation, the underlying physics is fundamentally different and therefore $\theta_{\text{CW}}$ and $\mu_{\text{eff}}$ do not retain the same meaning. Unlike their local moment counterparts, where only certain values of the effective moment, $\mu_{\text{eff}} = \sqrt{8C}$, are allowed due to the quantized values of the angular momenta quantum numbers, in itinerant magnets, atomic spin is no longer well-defined. Therefore, the fitted effective moment for an itinerant magnet, which depends on details of the electronic band structure, is not confined to take any specific values and is often smaller than the smallest local moments.

\subsection*{Phase transitions}
\label{PhaseTransitions}

The detection of a magnetic ordering or spin freezing transition via magnetic susceptibility is generally straightforward, as they are typically marked by a sharp discontinuity. The qualitative appearance of the discontinuity, its field dependence, and its response to field-cooled (FC) and zero-field-cooled (ZFC) conditions provides key insights that can be used to deduce the nature of the ordered or frozen state (see Box 1). However, in all but the simplest cases, determining the exact spin configuration requires additional experimentation -- specifically, neutron diffraction. The qualitative behaviors of various phase transitions that can be detected via susceptibility are described in Box 2 and representative examples are shown in Fig.~\ref{Transitons}.

Some magnetic materials may exhibit no magnetic ordering or freezing down to the lowest measurable temperatures. For example, materials with a small concentration of magnetic ions (also known as dilute magnets) may fall below the percolation threshold, which defines the density of magnetic ions required to obtain collective behavior depending on the geometry of the lattice, which is the case for the data set shown in Fig.~\ref{Transitons}(a). Another scenario through which ordering transitions are suppressed is magnetic frustration, as will be discussed in a later section. While these two scenarios can, at first blush, appear similar, the former will yield Curie-Weiss temperatures close to zero while the latter should not. A rare, but potentially confounding scenario can occur if your material orders above the measured temperature range, which is 400~K in typical set-ups. In this case your measurement will not conform with any of the behaviors described here since the paramagnetic regime is never accessed. 

\vspace{3mm}

\begin{tcolorbox}[breakable]
\subsection*{Box 2: Signatures of various types of phase transitions in magnetic susceptibility}

\begin{itemize}[leftmargin=*]

\item \textbf{Ferromagnets} order at the Curie temperature ($T_{\text{C}}$), which is marked by a divergent susceptibility due to the moments spontaneously aligning with the applied magnetic field. In ferromagnets with hysteresis, careful low-field measurements under FC and ZFC conditions will reveal a large splitting at temperatures below $T_{\text{C}}$. This is due to the random orientation of pinned ferromagnetic domains in the ZFC state whereas in FC conditions the sample will form a single domain aligned with the external field. Ferromagnetic transitions can also occur without splitting at large fields or in systems with no measurable hysteresis. In either case, the saturation value of the susceptibility in FC conditions should correspond to the full magnetic moment (note the useful relationship that 5585 emu per mol = 1 $\mu_{\text{B}}$ per f.u.) 

\item \textbf{Ferrimagnets} have a susceptibility that qualitatively resembles a ferromagnet, particularly in the vicinity of $T_{\text{C}}$. Ferrimagnetic transitions can be distinguished from ferromagnetic transitions by the saturation value of their susceptibility, which will be smaller than the full magnetic moment. If the two sublattices in a ferrimagnet have different ordering temperatures or if their ordered moments grow at different rates, one can see additional features beyond what is expected in a simple ferromagnet. For example, the susceptibility of a ferrimagnet may approach zero at the so-called compensation temperature, where the moments on the two sublattices precisely cancel one another out.

\item \textbf{Antiferromagnets} order at the N\'eel temperature ($T_{\text{N}}$), which is marked by a cusp in susceptibility. In textbook cases, the susceptibility will continue to decrease towards the lowest temperatures. However, in many real materials the susceptibility will continue to increase below the ordering transition due to paramagnetic impurities, often termed a Curie tail. Curie tails, which go as $1/T$, can also be observed in non-magnetically ordered materials. For a simple antiferromagnet, $T_{\text{N}}$ will be suppressed with increasing field. However, many antiferromagnets have complex temperature-field phase diagrams due to competing interactions.  

\item \textbf{Spin glasses} freeze at a temperature $T_{\text{f}}$ that is usually marked by a cusp in susceptibility closely resembling the cusps seen in antiferromagnets, when measured under ZFC conditions. However, when measured under FC conditions the susceptibility of a spin glass will be temperature independent below $T_{\text{f}}$, producing a large FC-ZFC splitting. The smoking gun signature for a spin glass transition is frequency dependent AC susceptibility measurements, where the transition should shift to higher temperatures with increasing field oscillation frequency. 

\item \textbf{Superconductors} have a zero resistance state giving rise to perfect diamagnetic screening and marked by an abrupt decrease in the susceptibility. In the superconducting state, the volume susceptibility is exactly $-1$ in SI units or $\sfrac{-1}{4\pi}$ in cgs units. However, one must take care to cool the sample through the superconducting transition in ZF conditions and measure the susceptibility in a suitably small field so as not to disrupt the superconducting state, typically 10 Oe or smaller. In real measurements, significant deviations from the perfect superconducting volume susceptibility are typical due to inhomogeneity or demagnetizing (shape) effects.

\end{itemize}

\end{tcolorbox}

\section*{Practical considerations} 
\label{practical}

With this brief synopsis on the theoretical underpinnings of magnetic susceptibility, we next move on to a topic of practical importance -- how to collect a high quality data set and, then, how to perform a quantitative analysis. Our focus here is measurements on polycrystalline or so-called ``powder'' samples, which are often more readily attainable. In polycrystalline measurements, rather than applying the magnetic field along one distinct crystallographic direction, all sample orientations are measured simultaneously and averaged over. Thereby, any information about anisotropy (directional dependence, see Box 1) is lost. In this section, we suggest some simple steps that we believe can help a newcomer to quickly assess their data set and begin to make sense of it.

\begin{figure} 
  \centering
  \includegraphics[width= 8.5cm]{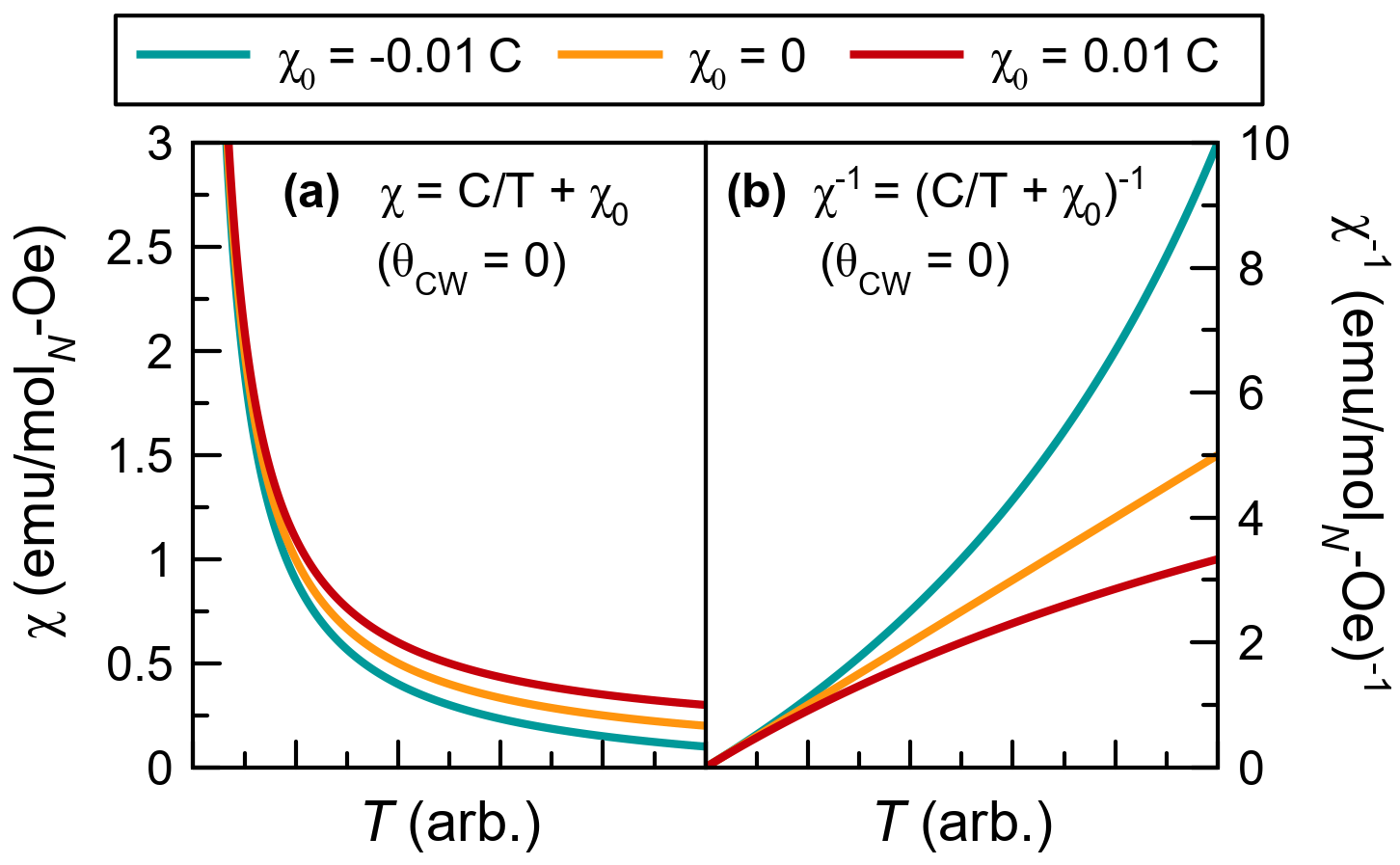}
  \caption{The effect of a small temperature independent contribution to the magnetic susceptibility, $\chi_0$, on Curie-Weiss behavior. The yellow curve shows the behavior or a material with $\chi_0 = 0$ while blue and red include, respectively, a small negative and a small positive $\chi_0$. \textbf{(a)} In the direct susceptibility, Curie-Weiss behavior is still evident for all three curves, \textbf{(b)} while in the inverse susceptibility a significant positive (red) or negative (blue) curvature is observed that must be accounted for (see Eqn.~\ref{chi0eqn}) in order to fit the data}.
  \label{chi0fig}
\end{figure}

\begin{figure*}
  \centering
  \includegraphics[width=18cm]{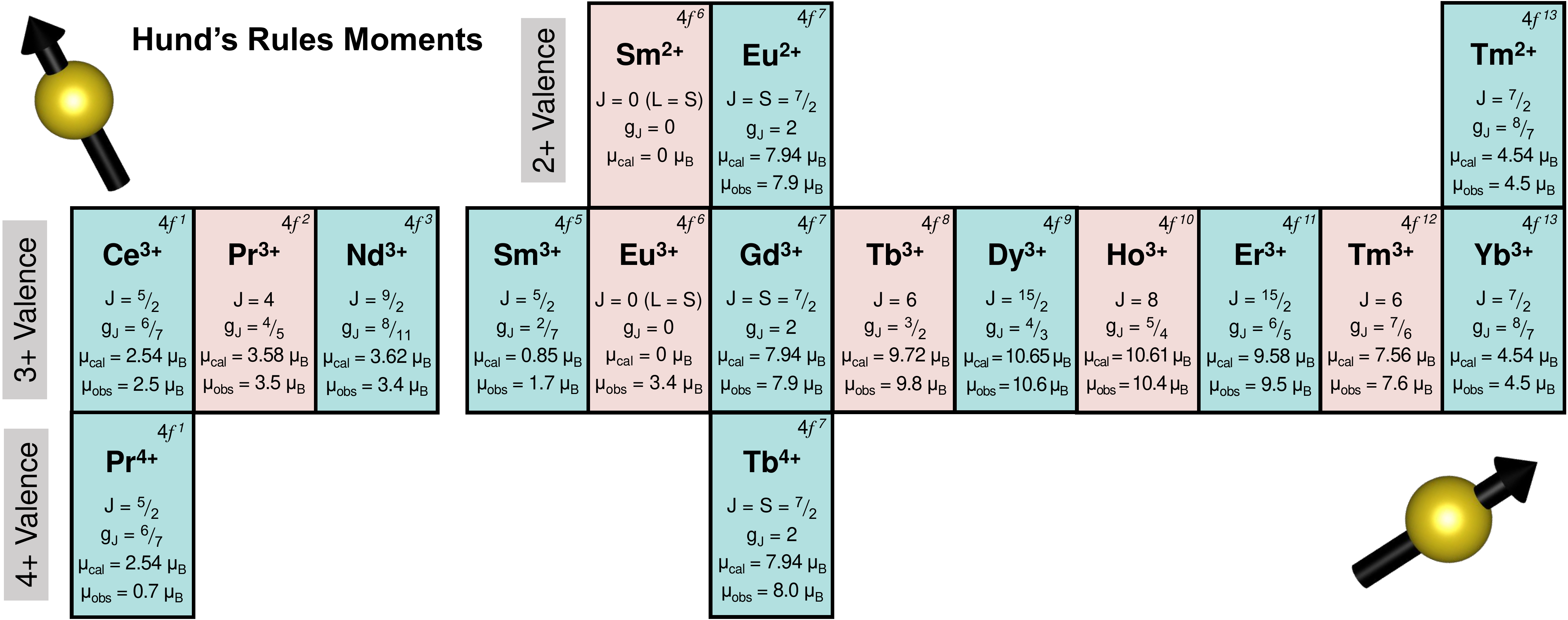}
  \caption{Schematic illustration of the magnetic properties of rare earth ions arranged according to their number of $f$ electrons in their divalent (top row), trivalent (middle row) and tetravalent (bottom row) states. Each entry includes the Hund's rules total angular momentum $J$, the Land\'e g-factor $g_J$, the calculated paramagnetic moment $\mu_{\text{cal}} = g_J\sqrt{J(J+1)}$, and the experimentally observed paramagnetic moment $\mu_{\text{eff}}$. Excellent agreement between the calculated and observed moments is found for almost all rare earths. Poor agreement for Sm$^{3+}$, Eu$^{3+}$, and Pr$^{4+}$ is due to the proximity of the next highest spin-orbit manifold. The blue shaded ions have an odd number of electrons and are therefore Kramers ions, which are guaranteed to have at least a ground state doublet while the pink shaded ions are non-Kramers ions, where a non-magnetic singlet ground state is allowed by symmetry. Non-magnetic Ce$^{4+}$ and Yb$^{2+}$ are not included here.}
  \label{RareEarth_mu}
\end{figure*}

\begin{enumerate}[leftmargin=*]

\item Accurately record the mass of your sample prior to your measurement. For best results, use a four place balance and measure in triplicate. 

\item Optimize your measurement parameters. This includes mounting your sample in a way minimizes the background contribution (for polycrystalline samples, we suggest an inverted gel capsule), and choosing an appropriate mass of sample and applied field magnitude. For best results, consult your instrument's user manual. Increasing the sample mass and applied field magnitude will both have the effect of increasing your overall signal to noise. However, be aware that the assumed linear relationship between $M$ and $H$ can break down at high fields.

\item Convert the measured magnetization into molar susceptibility using the following formula
\begin{equation}
\chi_{\text{mol}} [\textrm{emu\,mol$^{-1}_{N}$\,Oe$^{-1}$}] = \frac{M [\textrm{emu}]}{H [\textrm{Oe}]} \left( \frac{M [\textrm{g\,mol$^{-1}$}]}{m [\textrm{g}] \cdot n} \right)
\end{equation}
where $M$ confusingly denotes the magnetization, in the first term, and the molar mass, in the second term, $H$ is the applied field, $m$ is the sample mass, and $n$ is the number of magnetic ions, $N$, per formula unit. Units are indicated for each term in the square brackets. In cases where the number of magnetic ions is unknown, you can use $n = 1$ to obtain molar susceptibility per formula unit. 

\item Use this data to generate a plot of $\chi_{\text{mol}}$ vs. temperature, $T$. Consider what contributions to susceptibility you expect to be present in your material (making use of the previous section). The steps that follow are specifically for systems in which a Curie-Weiss (local moment) contribution is expected. In those cases, at temperatures above any magnetic ordering or freezing transition, you should see a susceptibility that rapidly increases with decreasing temperature. 

\item Plot $\chi^{-1}_{\text{mol}}$ vs. $T$. If a linear regime is observed, fit that data to the Curie-Weiss equation
\begin{equation} \label{inverse}
\chi^{-1} = \frac{T-\theta_{\text{CW}}}{C} = \frac{T}{C}-\frac{\theta_{\text{CW}}}{C} 
\end{equation}
yielding a linear relationship in which the slope is $\sfrac{1}{C}$, the $y$-intercept is $\sfrac{\theta_{\text{CW}}}{C}$, and the $x$-intercept is $\theta_{\text{CW}}$.
Using Eqn.~\ref{effective}, calculate the effective moment, $\mu_{\text{eff}}$, and compare that value with what is expected for your magnetic ion using Eqn.~\ref{mucalc} (more details on this are provided in the respective sections on $4f$ rare earth magnets and $3d$ transition metals).

\item All fitting should be performed with $\chi^{-1}$ vs. $T$ and not $\chi$ vs. $T$. This is because in $\chi$ vs. $T$, chi-squared minimization fitting routines will overly weight the lowest temperature data points where the susceptibility is largest but where the expected adherence to Curie-Weiss behavior is worst. Also, temperature independent contributions to the susceptibility can go undetected in $\chi$ vs. $T$ plots but not in $\chi^{-1}$ vs. $T$ (see Fig.~\ref{chi0fig}). The fitting range cutoff in $\chi^{-1}$ vs. $T$ should be chosen based on where visible deviations from Curie-Weiss behavior are observed.

\item If your plot of $\chi^{-1}_{\text{mol}}$ vs. $T$ shows positive or negative curvature, rather than a strictly linear trend, the likely origin is a temperature independent contribution to susceptibility, $\chi_0$. As shown in Fig.~\ref{chi0fig}, even a relatively small $\chi_0$ can result in significant curvature when the data is plotted as $\chi^{-1}$. This temperature independent term can have many origins including: core diamagnetism (both from the sample or from the sample holder), Pauli paramagnetism, or van Vleck paramagnetism. If the curvature is positive (red curve in Fig.~\ref{chi0fig}) this indicates a positive  $\chi_0$ while if the curvature is negative (blue curve) this indicates a negative $\chi_0$. In this case a modified form of the Curie-Weiss law can be applied.
\begin{equation} \label{chi0eqn}
\chi = \frac{C}{T-\theta_{\text{CW}}} + \chi_0, \textrm{~~}
\chi^{-1} = \frac{T-\theta_{\text{CW}}}{\chi_0 \cdot (T-\theta_{\text{CW}})+C}
\end{equation}
At this stage, it is wise to verify that the magnitude and sign of $\chi_0$ is physically consistent with its expected origin.

\end{enumerate}

While these generic steps will work in many cases, there are times that they will fail to yield meaningful results. It is equally important to avoid misapplying the Curie-Weiss law -- an equation that was described by Van Vleck as ``the most overworked formula in the history of paramagnetism~\cite{van1973chi}.'' If there is no physical justification for local magnetic moments in the sample you are studying, then the results obtained from a Curie-Weiss fitting will not be valid, even if the equation fits. Even in cases where local moments are present, there are other factors that can lead to the breakdown of Curie-Weiss behavior, such as low-dimensionality, high-spin to low-spin crossovers, and intermediate valence. Each of these gives rise to a characteristic temperature-dependence in susceptibility that disrupts conventional Curie-Weiss behavior.  In the sections that follow, we present specific case studies among $3d$ transition metal and $4f$ rare earth based systems. We consider systems that both adhere to conventional Curie-Weiss behavior as well as exceptions to the rule.

\section*{4\lowercase{f} Rare Earth Magnets}
\label{4f}

In the $4f$ rare earth block there is a clear hierarchy of energy scales that guides the interpretation of magnetic susceptibility data. Due to their large mass, spin-orbit coupling, which scales as $Z^4$, dominates over all other magnetic energy scales. As a result, magnetic rare earth elements are the canonical Hund's rules ions. The partially filled $4f$ orbitals are highly localized and shielded from their neighboring ions by the more spatially extended, fully occupied $5s$ and $5p$ orbitals and, as a result, the crystal electric field is significantly weakened. These highly localized $4f$ orbitals also have minimal direct or indirect orbital overlap. Consequently, in insulating rare earth magnets, exchange interactions (see Box 1) tend to be very weak and magnetic ordering temperatures of 1 K or lower are typical. In metallic rare earth magnets, while the $4f$ moments themselves generally retain a local character, the interactions can be enhanced due to conduction electron mediated exchange via the RKKY mechanism~\cite{ruderman1954indirect,kasuya1956theory,yosida1957magnetic}, leading to magnetic ordering temperatures on the order of 10 K.

Rare earths (with a few notable exceptions) are almost always found in their trivalent state~~\cite{strange1999understanding}, $R^{3+}$, with a total angular momentum, $J$, accurately estimated using Hund's rules. This angular momentum defines the spin orbit ground state manifold, which has a degeneracy of $2J+1$ states. Higher order spin-orbit manifolds are split by an amount proportional to the atomic spin-orbit coupling, which for rare earths is typically 1 eV ($\approx 10^4$ K) or higher. Therefore, at the temperatures relevant to a typical susceptibility measurement, one can ignore them altogether. Inspection of Figure~\ref{RareEarth_mu} shows that the calculated Hund's rules moments ($\mu_{\text{cal}}$) are in excellent agreement with the observed values ($\mu_{\text{obs}}$) for the majority of rare earth ions. Next, the crystal field lifts the degeneracy of the $2J+1$ states, inducing crystal field splittings between 10 and 100 meV. This crystal field splitting can introduce intense spin anisotropy. This anisotropy, which often manifests as directional dependence in single crystal susceptibility measurements, is most pronounced at low temperatures but in some cases can extend significantly beyond room temperature. In measurements of polycrystalline samples this information is lost due to directional averaging. In the case of ions with an odd number of total electrons, indicated in blue in Figure~\ref{RareEarth_mu}, a magnetic (doublet) ground state is guaranteed by Kramers theorem, while those with even electron counts, shown in pink, can have non-magnetic singlet ground states depending on the specific nature of the crystal field splitting.

\begin{figure}
  \centering
  \includegraphics[width=8.5cm]{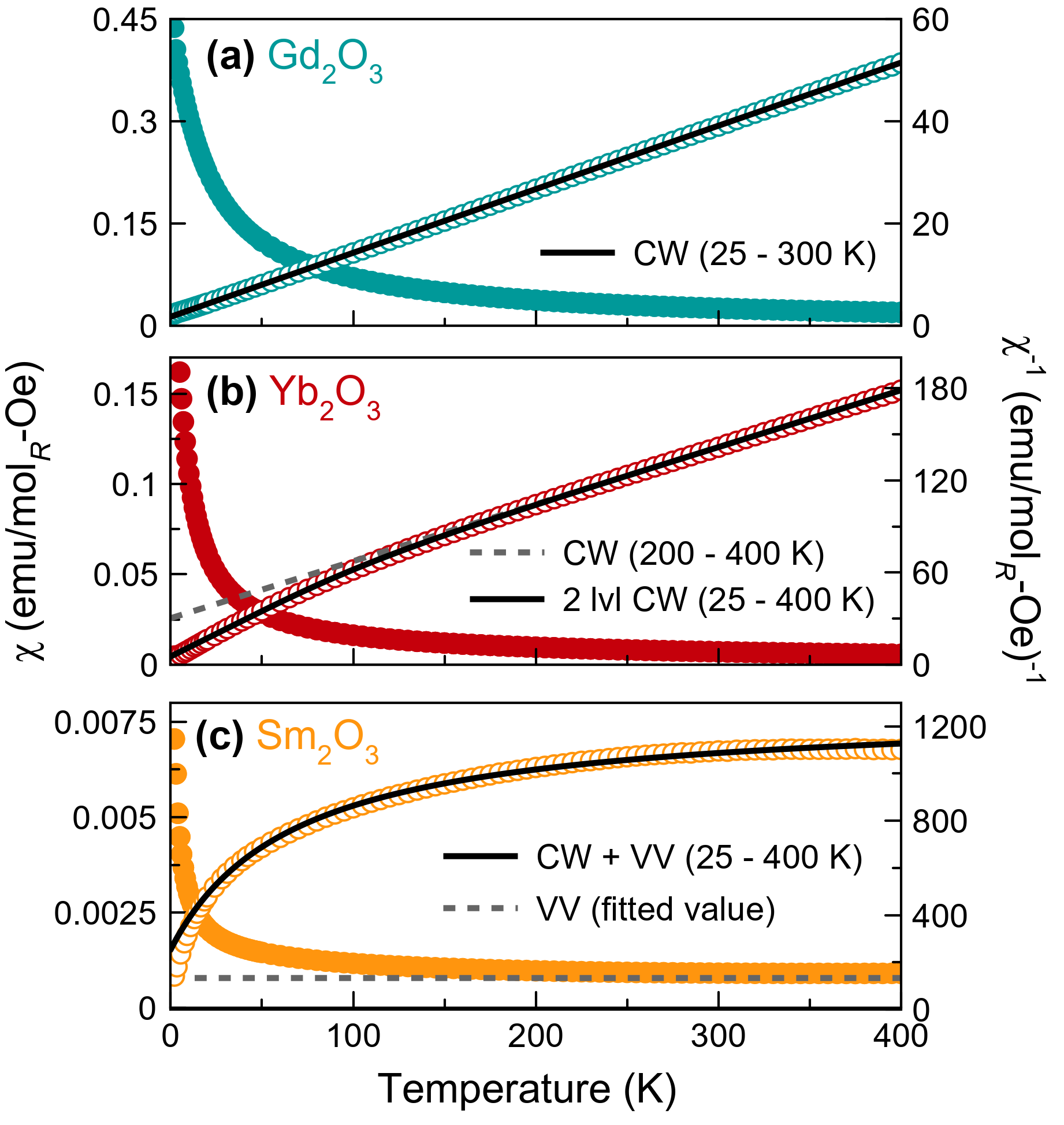}
  \caption{Representative examples of Curie-Weiss behavior in rare earth magnets. The magnetic susceptibility (left axis, filled symbols) and inverse magnetic susceptibility (right axis, open symbols) normalized per mole of rare earth for \textbf{(a)} Gd$_2$O$_3$, which is well-fitted by a standard Curie-Weiss (CW) law, \textbf{(b)} Yb$_2$O$_3$, which is well-fitted by a modified Curie-Weiss law that incorporates an excited crystal field, and \textbf{(c)} Sm$_2$O$_3$, which is well-fitted by a Curie-Weiss law plus a van Vleck (VV) contribution, all measured in an applied field of $H = 1000$~Oe.}
  \label{RareEarth_CW}
\end{figure}

The question we now come to is, over what temperature range should the Curie-Weiss law be applied in rare earth magnets? The exact temperature range is, of course, material dependent but for most rare earth magnets reasonable adherence to Curie-Weiss behavior is obtained in the temperature range of 200 to 400 K. The reason for the breakdown of Curie-Weiss behavior at lower temperature is two-fold. First, as temperature decreases, the excited crystal field levels become thermally depopulated as the material enters its crystal field ground state. Depending upon which $m_J$ basis states form the crystal field ground state, the magnetic moment can significantly decrease from the Hund's rules value. Secondly, at lower temperatures, when the energy scale of the magnetic interactions becomes comparable with the thermal energy, the system can no longer be treated as an uncorrelated paramagnet. 

A noteworthy exception occurs for an exactly half-filled $4f$ electron shell, as is the case for Eu$^{2+}$, Gd$^{3+}$, and Tb$^{4+}$ (Fig.~\ref{RareEarth_mu}). For these ions, there is no orbital ($L$) component to the total angular momentum. Consequently, the crystal field leaves the Hund's rules manifold nearly degenerate and Curie-Weiss behavior is observed to much lower temperatures, until interactions set in, as can be seen in the susceptibility of Gd$_2$O$_3$ shown in Fig.~\ref{RareEarth_CW}(a). The data can be well fit by a Curie-Weiss Eqn.~\ref{inverse} from 400~K all the way down to 25~K. This fit yields an effective moment of $\mu_{\text{eff}}=7.96$~$\mu_{\text{B}}$, which is in excellent agreement with the Hund's rules moment of $\mu_{\text{calc}}=7.94$~$\mu_{\text{B}}$ for Gd$^{3+}$, while $\theta_{\text{CW}}=-14$~K, which is reasonable given that this material is known to order antiferromagnetically at $T_{\text{N}} = 3.8$~K~\cite{hill1983specific}. 

In other rare earths, the value of $\theta_{\text{CW}}$, must be treated with caution. Except in the case of the exactly half-filled $4f$ electron shell described above, a high temperature fitting of $\theta_{\text{CW}}$ will include contributions from thermally populated crystal field levels that are not present at the low temperatures where interactions become relevant. Therefore, $\theta_{\text{CW}}$ in most rare earth magnets will invariably overestimate the strength of the interactions. For example, a Curie-Weiss fit to the susceptibility of Yb$_2$O$_3$ between 200 and 400 K (Fig~\ref{RareEarth_CW}(b)) gives an effective moment of $\mu_{\text{eff}}=4.64$~$\mu_{\text{B}}$, in good agreement with $\mu_{\text{calc}}=4.54$~$\mu_{\text{B}}$ for Yb$^{3+}$ (Fig.~\ref{RareEarth_mu}), but an unphysically large $\theta_{\text{CW}}=-81$~K. A rigorous treatment of this problem would require an experimental determination of the material's full crystal field scheme, as can be achieved with inelastic neutron scattering or Raman spectroscopy. One can, however, approximate the effect of excited crystal field levels using the following two-level equation~\cite{mitric1997x,besara2014single},
\begin{equation}
\label{Eqn:ExpandedCW}
\chi^{-1} = 8 \cdot (T-\theta_{\text{CW}}) \cdot \left( \frac{\mu_{\text{eff,0}}^2 + \mu_{\text{eff,1}}^2 \cdot e^{- \frac{E_1}{k_{\text{B}} T}}}{1 + e^{- \frac{E_1}{k_{\text{B}} T}} } \right)
\end{equation}
where $E_1$ is the energy splitting to the first excited crystal field level, with an effective moments of $\mu_{\text{eff,1}}$, while $\mu_{\text{eff,0}}$ is the effective moment of the crystal field ground state. Applying this model to Yb$_2$O$_3$ extends the fits to significantly lower temperature, 25 to 400 K, and yields a more physical value of $\theta_{\text{CW}}=-9$~K and a ground state moment of $\mu_{\text{eff,0}} = 3.67$~$\mu_{\text{B}}$, consistent with a crystal field ground state made-up of largely $m_J = \sfrac{5}{2}$. This equation can be extended to include a second excited crystal field level, but this leads to a highly under constrained fit, with too many adjustable parameters to be reasonably determined by a 1-dimensional data set.

In relatively few rare earth magnets, straightforward Curie-Weiss behavior is not observed over any temperature window. There are a few ion specific factors that can lead to this effect. One is an appreciable temperature independent van Vleck contribution to susceptibility, which frequently occurs for Sm$^{3+}$ and Eu$^{3+}$ compounds, due to their smaller splitting to the first excited spin-orbit manifolds. The telltale sign for this effect is pronounced curvature in the inverse susceptibility, $\chi^{-1}$, as shown for Sm$_2$O$_3$ in Fig.~\ref{RareEarth_CW}(c). A good fit of the data can be obtained by including a temperature independent term, as given by given by Eqn.~\ref{chi0eqn}, which gives a van Vleck contribution of $\chi_{\text{vv}} = 7.9 \times 10^{-4}$~emu\,mol$^{-1}_{\text{Sm}}$, indicated by the grey dashed line, $\theta_{\text{CW}}=-13$~K, and $\mu_{\text{eff}}=0.58$~$\mu_{\text{B}}$. Another factor that can produce significant curvature in $\chi^{-1}$ in metallic rare earth magnets is a large Pauli contribution to the susceptibility, $\chi_{\text{P}}$. In this case one can fit to the same modified version of the Curie-Weiss law. 

Finally, in the special cases of Ce$^{3+}$ ($4f^1$) and Yb$^{3+}$ ($4f^{13}$), which are both a single electron from a closed shell configuration, there is an instability towards the 4+ and 2+ valence, respectively. As a result, some unique behaviors can arise. Some metallic rare earth compounds will exhibit an intermediate valence due to a near degeneracy of the two valence configurations~\cite{varma1976mixed}, which produces a characteristic broad hump in susceptibility and a disruption of Curie-Weiss behavior~\cite{sales1975susceptibility}. In systems in which the rare earth ion occupies multiple crystallographic sites, a mixture of valences is also possible such that some sites are magnetic and some are non-magnetic, in which case Curie-Weiss behavior should be retained.

\begin{figure*}
  \centering
  \includegraphics[width=18cm]{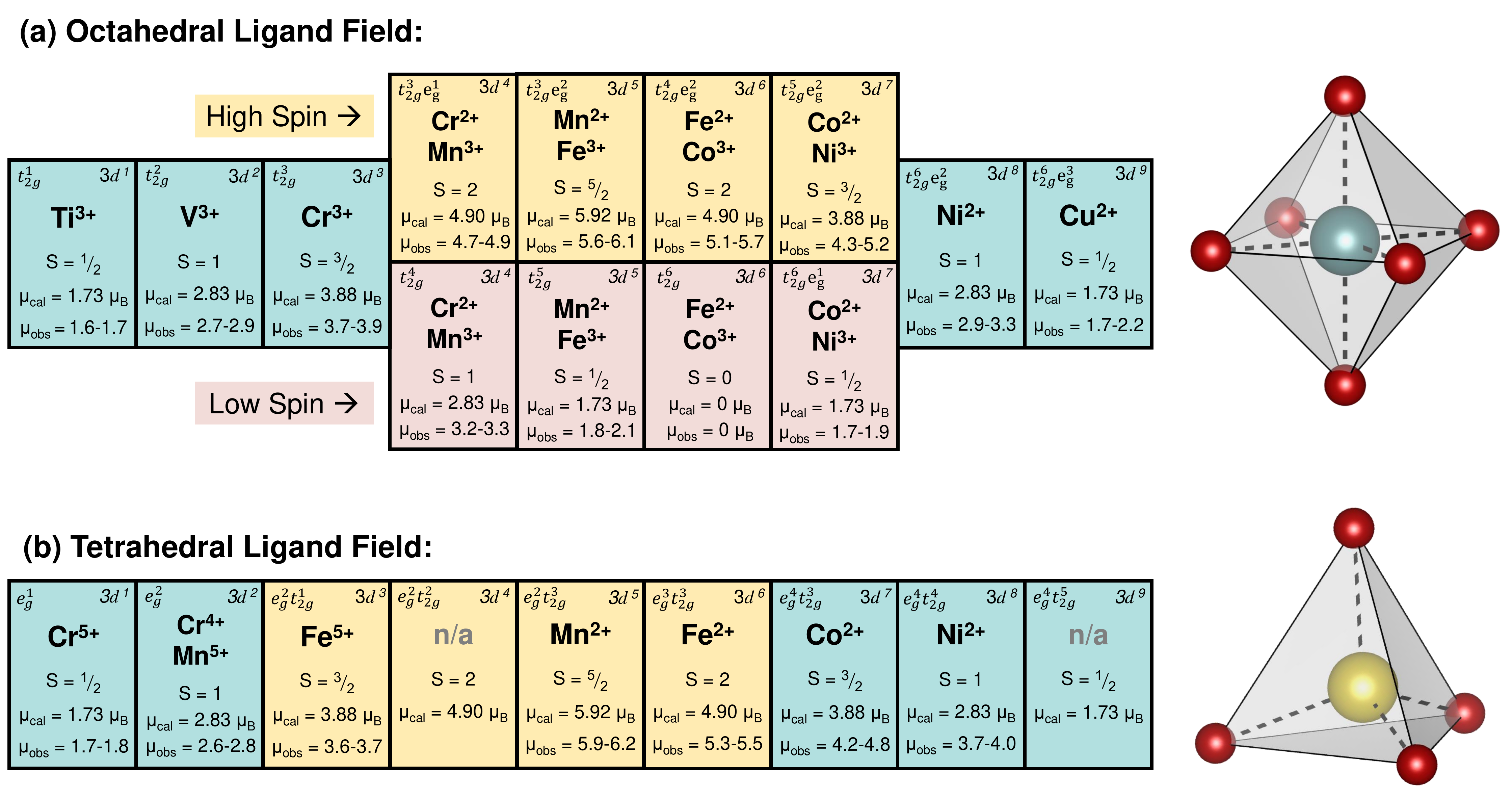}
    \caption{Schematic illustration of the magnetic properties of various $3d$ transition metals arranged according to their number of $d$ electrons and in their common valence states in \textbf{(a)} octahedral and \textbf{(b)} tetrahedral ligand fields. Due to quenching of orbital angular momentum, their calculated moments, $\mu_{\text{cal}} = g\sqrt{S(S+1)}$, are best estimated from the spin-only, $S$, value with an isotropic Land\'e g-factor, $g=2$, giving good agreement with the observed moments, $\mu_{\text{obs}}$. In the octahedral case, the crystal field ground state is formed by the three degenerate $t_{\text{2g}}$ states with a crystal field splitting to the two higher energy $e_{\text{g}}$ states, while the opposite pattern is observed for the tetrahedral case. When there are between four ($3d^4$) or seven ($3d^7$) electrons in the $3d$-orbitals, either high-spin (yellow) or low-spin (pink) states can be observed for the octahedral case, with the overall crystal field splitting determining which is energetically preferred. The smaller crystal field splitting in the tetrahedral case means that high-spin states are always obtained, with very few exceptions. Observed moments that exceed the calculated moment, such as for Co$^{2+}$, imply an intermediate spin-orbit coupling and an incomplete quenching of orbital angular momentum.}
  \label{3d_periodic}
\end{figure*}

\section*{3\lowercase{d} Transition Metal Magnets}
\label{3d}

Moving into the $3d$ transition metal block there are a number of significant differences from the $4f$ case that one must pay attention to when interpreting susceptibility data. First we will consider the critical difference in the nature of the partially filled electron orbitals themselves. Whereas $4f$ electron orbitals are spatially localized and are therefore not significantly involved in either chemical bonding or in electrical conduction, partially filled $3d$ orbitals are significantly involved in both. This complicates matters, particularly in the case of metallic $3d$ transition metal compounds, where electronic and magnetic degrees of freedom may be strongly intertwined. In some cases, the magnetism can be quenched altogether giving rise to a paramagnetic metal. However, in cases where the magnetism persists, the moments will have a  mixed local and itinerant character that even in the case of the seemingly simple elemental metals (Cr, Fe, Co, Ni) remains a challenge to theoretically describe~\cite{eich2017band}. We will therefore focus our remarks here on insulating materials with unpaired $3d$ electrons. 

Having narrowed our scope to insulators we next consider the reorganization of energy scales, which in turn modifies the magnetic character of the $3d$ transition metal ions. Since the partially filled $d$ electron orbitals are also the most spatially extended, they are the ones that participate in chemical bonding. As a result, their direct overlap with the ions in their local environment (ligands) is larger and the energy scale of the crystal electric field is substantially enhanced. In fact, for $3d$ transition metal compounds, the crystal field is the largest magnetic energy scale, with splittings larger than 1~eV being common, depending on the ligand. The filling of the $3d$ electron orbitals is strongly determined by the symmetry of the local environment, as illustrated in Fig.~\ref{3d_periodic} for octahedral and tetrahedral ligand fields, which induce a splitting between the $t_{\text{2g}}$ states ($d_{xy}$, $d_{xz}$, $d_{yz}$) and the $e_{\text{g}}$ states ($d_{z^2}$, $d_{x^2-y^2}$). In the octahedral case, depending on the magnitude of the splitting, either high-spin or low-spin states can be found, as indicated by the yellow and pink shading in Fig.~\ref{3d_periodic}, while in the tetrahedral case the energy splitting is smaller and only high-spin states are commonly observed. It is important to emphasize that while octahedral and tetrahedral ligand fields are the most common for $3d$ compounds they are by no means the only local environments you can encounter, which will in turn modify the crystal field splittings between $3d$ orbitals and the number of unpaired $d$ electrons. For a more detailed overview of ligand field theory we refer the reader to the classic texts on this topic~\cite{griffith1964theory,figgis1999ligand}. 

The modified energy scale also changes the nature of the $3d$ magnetic moment in-and-of-itself due to the breakdown of Hund's rules. First, as compared to the $4f$ rare earths described previously, spin orbit coupling (which scales as $Z^4$) no longer dominates and, in fact, can largely be ignored. Further, as a result of the enhanced Coulomb interactions with surrounding ions, it is too energetically costly to transform one $3d$ electron orbital into another via rotation, meaning that the time-averaged magnitude of the orbital moment approaches zero, which is termed ``quenching'' of orbital angular momentum (Kittel provides a detailed quantum mechanical treatment~\cite{kittel}). In $3d$ magnets, one finds that the magnetic moment is very well approximated by the spin-only moment, which means that the problem is essentially reduced to one of counting unpaired electrons. The spin-only moment can be calculated using Eqn.~\ref{mucalc} by replacing $J$ with $S$ and where $g=2$, reflecting the isotropic nature of the moment, giving
\begin{equation} \label{mucalc3d}
\mu_{\text{cal}} = 2 \sqrt{S(S+1)} \; \mu_{\text{B}}.
\end{equation}
Examination of the calculated spin-only moments, $\mu_{\text{cal}}$, and the commonly observed moments, $\mu_{\text{obs}}$, in Fig.~\ref{3d_periodic} generally shows very good agreement, particularly for the ions with the smallest mass (spin-orbit coupling). Finally, larger orbital overlap also leads to an enhancement of the superexchange interactions, with $3d$ transition metal oxides commonly ordering at temperatures on the order of 100~K, with a small number of compounds even exceeding 1000~K.

\begin{figure}
  \centering
  \includegraphics[width=8.5cm]{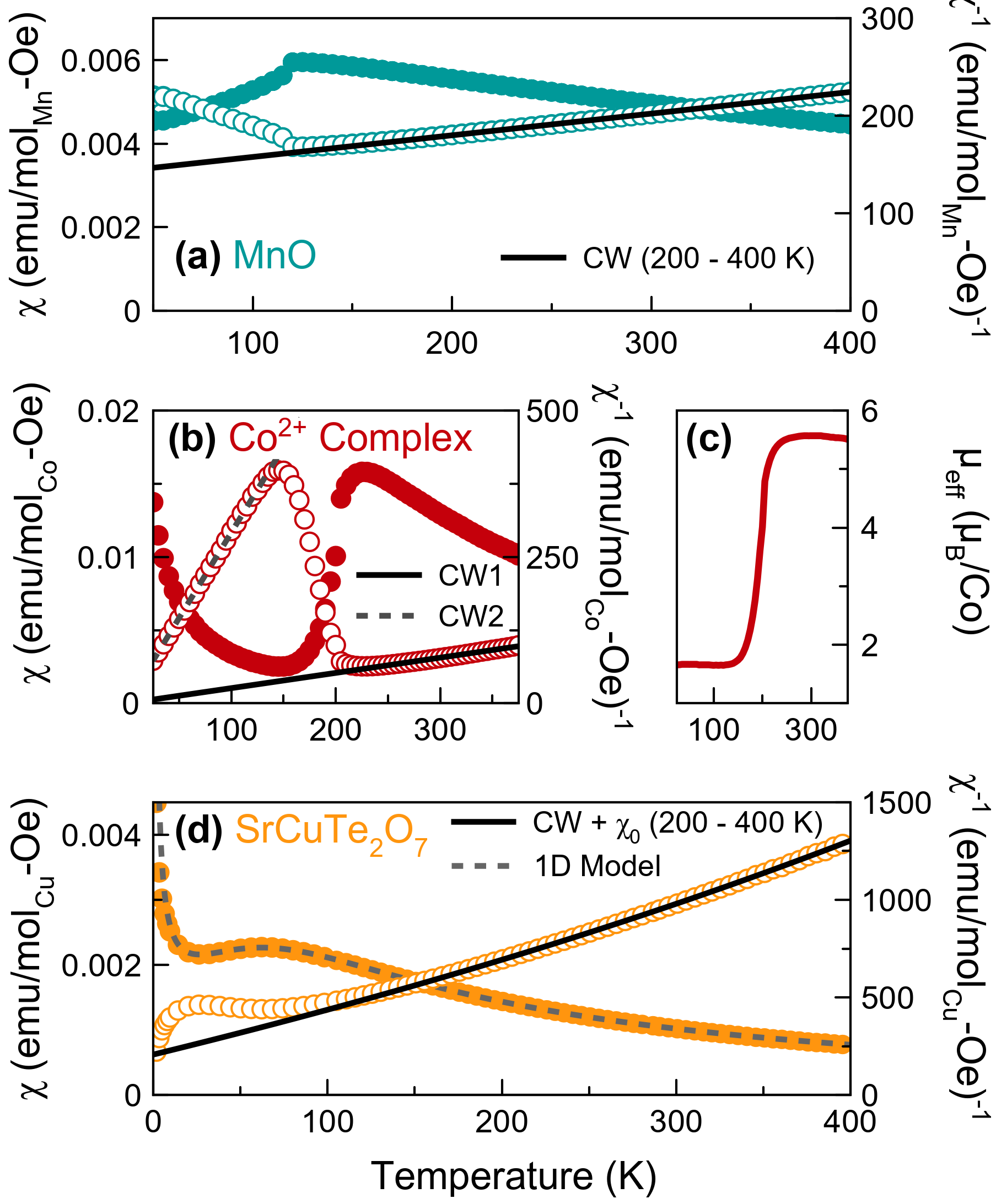}
  \caption{Representative examples of Curie-Weiss behavior in $3d$ transition metal magnets. The magnetic susceptibility (left axis, filled symbols) and inverse magnetic susceptibility (right axis, open symbols) normalized per mole of transition metal for \textbf{(a)} MnO measured in a field of $H=1000$~Oe, which is well-fitted by a standard Curie-Weiss (CW) law, \textbf{(b)} Reproduced data for Co(tmeda)(3,5-DBQ)$_2 \cdot $0.5C$_6$H$_6$ measured in a field of $H=5000$~Oe~\cite{liang2017charge}, which is an organic Co$^{2+}$ complex that undergoes a crossover from high-spin to low-spin just below 200 K with two distinct Curie-Weiss regimes. Panel \textbf{(c)} shows the temperature dependence of $\mu_{\text{eff}} = \sqrt{8 \chi T}$ as a function of temperature, which is the conventional way of presenting susceptibility data for a spin crossover complexes, and \textbf{(d)} SrCuTe$_2$O$_7$ measured in a field of $H=10,000$~Oe, which is well-fitted by a Curie-Weiss law at temperatures above 200 K but exhibits an anomaly characteristic of 1-dimensional interactions at lower temperature.}
  \label{3dtransition_CW}
\end{figure}

This enhancement of exchange interactions is apparent in the magnetic susceptibility data of MnO, shown in Fig.~\ref{3dtransition_CW}(a), which exhibits a pronounced cusp at $T_{\text{N}} = 120$~K, marking an antiferromagnetic ordering transition. Plotting the inverse susceptibility reveals Curie-Weiss like behavior that extends from 400 K to just above the ordering transition. Applying Eqn.~\ref{mucalc3d} between 200 and 400~K gives a fitted moment of $\mu_{\text{eff}}=6.00$~$\mu_{\text{B}}$, which agrees very well with the expected high-spin moment for Mn$^{2+}$ in an octahedral ligand field with five unpaired $d$ electrons (Fig.~\ref{3d_periodic}(a)). The Curie-Weiss temperature is calculated to be $\theta_{\text{CW}}= -610$~K, which is five-fold larger than the magnetic ordering temperature, but still in the typical range for antiferromagnets as described previously. 

Another important distinction between $4f$ and $3d$ magnetism is that in the former case, we expect deviations from Curie-Weiss behavior due to the changing magnetic moment associated with the gradual thermal de-population of excited crystal field states. In contrast, magnetic moments are most often independent of temperature in a $3d$ transition metal compound, as is the case in the previous example of MnO. However, there are exceptions to this rule in cases where the energy of high-spin and low-spin configurations are similar, as can occur for ions with between four and seven unpaired $3d$ electrons in octahedral ligand fields (Fig.~\ref{3d_periodic}(a)). In such cases, a temperature-induced crossover between high-spin and low-spin configurations can occur, which is associated with decreased lattice vibrations. Such crossovers are most frequently observed in compounds with organic ligands where inter-ion magnetic interactions are approaching non-existent. In contrast to the gentle curvature associated with thermal depopulation of rare earth crystal field levels, these spin crossovers are typically abrupt, giving rise to a sharp decrease in magnetic moment, such as the one seen for Co(tmeda)(3,5-DBQ)$_2 \cdot $0.5C$_6$H$_6$, an organic Co$^{2+}$ compound, just below 200~K in Fig.~\ref{3dtransition_CW}(b)~\cite{liang2017charge}. This crossover can be appreciated most clearly by plotting $\mu_{\text{eff}} = \sqrt{8 \chi T}$ as a function of temperature as shown in Fig.~\ref{3dtransition_CW}(c), under the assumption that correlations are negligible ($\theta_{\text{CW}}=0$). In the high temperature regime, the effective moment is $\mu_{\text{eff}}=5.5$~$\mu_{\text{B}}$, in good agreement with the high-spin moment for Co$^{2+}$, which has a moderate orbital contribution. In the low temperature limit the moment goes to $\mu_{\text{eff}}=1.7$~$\mu_{\text{B}}$, the expected moment for low-spin Co$^{2+}$ (Fig.~\ref{3d_periodic}(a)).

Another factor that can lead to the breakdown of Curie-Weiss behavior is low dimensional magnetism, and in particular quasi-1-dimensional magnetism. This type of behavior is observed when the magnetic exchange interactions are dominant along just one spatial direction, such that interactions in the other spatial directions can be largely neglected. Such materials, which are most commonly based on $3d$-transition metals, can usually be identified from their crystallography as the 1-dimensional nature of the interactions is underpinned by the atomic arrangement of the magnetic ions into chain-like motifs. In the high temperature paramagnetic limit, where thermal fluctuations are dominant, a quasi-1-dimensional magnet should exhibit the typical Curie Weiss susceptibility according to its spin state and local environment (Fig.~\ref{3d_periodic}). However, in the lower temperature limit where the energy scale of spin-spin interactions become comparable with temperature, magnetic order is suppressed by the low-dimensionality and a characteristic broad hump is observed in susceptibility. This can be seen in the susceptibility of SrCuTe$_2$O$_7$, a material with zigzag chains, in Rig.~\ref{3dtransition_CW}(d). This material has Curie-Weiss like susceptibility that holds to approximately 100~K. A number of theoretical works studying 1-dimensional chain models have reproduced this characteristic broad maxima in the susceptibility~\cite{bonner1964linear,eggert1994susceptibility}, which can be straightforwardly fit in experimental data using a polynomial approximation~\cite{feyerherm2000magnetic} of the exact solution~\cite{eggert1994susceptibility}. The one shown for SrCuTe$_2$O$_7$ gives an exchange coupling strength of $J = 108$~K, commensurate with where the observed breakdown of Curie-Weiss behavior occurs, and an effective moment of $1.82$~$\mu_{\text{B}}$, consistent with the expected moment for Cu$^{2+}$. The various phenomena of low dimensional magnetism are covered in-depth by other sources \cite{giamarchi2003quantum,vasiliev2018milestones} and Landee and Turnbull give a detailed treatment of the magnetic susceptibility of low-dimensional magnets~\cite{landee2014gentle}.

\section*{$4d$ and $5d$ Transition Metal Magnets}
\label{4d5d}

Magnetic materials based on magnetic $4d$ and $5d$ transition metals are relatively fewer than their $3d$ and $4f$ counterparts. The partially filled $4d$ and $5d$ orbitals are even more spatially extended than in the $3d$ case, such that orbital overlap with neighboring ligands increases and thus the crystal field is very large. Meanwhile the on-site (Hubbard) repulsion for doubly occupying a given orbital is reduced. As a result of these two periodic trends, a large fraction of $4d$ and $5d$ materials are metals without localized magnetic moments. For example, not a single elemental metal in the $4d$ or $5d$ block is magnetically ordered, in contrast to the magnetic elemental metals that make up nearly half the $3d$ block. Likewise, magnetism (besides Pauli paramagnetism) in intermetallics based on $4d$ and $5d$ transition metals is almost unheard of~\cite{misawa19861}. In the smaller fraction of insulating $4d$ and $5d$ materials, unpaired valence electrons can still give rise to localized magnetic moments that will behave according to the Curie-Weiss law in the paramagnetic regime. This scenario is typically borne out in systems where inter-ion exchange is especially weak, as in molecular magnets~\cite{wang2011molecular} or in crystalline solids with particularly large spacings between neighboring $4d$ or $5d$ ions, such as double perovskites and related structures~\cite{greedan2011magnetic}. Insulating states can also be found in materials where the chemical bonding is strongly ionic, such as halides, or in materials where spin-orbit coupling is so intense as to produce a correlated Mott insulating state~\cite{witczak2014correlated}.

There are fewer hard and fast rules that can guide the interpretation of magnetic susceptibility data in the $4d$ and $5d$ block. These materials tend to exist in a realm that is intermediate to the quenched orbital angular momentum of $3d$ transition metals and the Hund's rules adherence of $4f$ rare earth ions. Spin orbit coupling is too large to be neglected but may not be so large as to generate a full orbital contribution to the moment (Ma \emph{et al.} provide calculated ion specific spin-orbit coupling constants~\cite{ma2014systematic}). Due to the stronger crystal field and weaker on-site repulsion, $4d$ and $5d$ ions are almost exclusively found in low spin electron configurations, regardless of the specific ligand. As a result, there is less variability in which ions can even be magnetic in the first place. Among the $4d$ block, only Mo, Ru, and Rh are commonly observed to be magnetic, while among the $5d$ block, local moment magnetism is only observed for Re, Os, and Ir. Further complicating matters is that due to their larger ionic radii, $4d$ and $5d$ ions are more commonly encountered in lower symmetry local environments than their $3d$ counterparts, meaning that the crystal field splitting can be more complex than the simple octahedral or tetrahedral cases described previously. Spin-orbit coupling can further split these crystal field states such that, in some cases, it is difficult to predict the paramagnetic moment a priori.

When Curie-Weiss behavior is observed in $4d$ and $5d$ magnets, it can sometimes be used to evaluate the strength of the spin-orbit coupling. Assuming that one has a reasonable grasp of the likely valence state of the $4d$ or $5d$ ion in question (say in the case of an oxide or halide where the valences of all other ions is known), then one can compare the size of the fitted paramagnetic moment with the expected value in the two limiting cases: the spin only moment and the full Hund’s rules moment. However, in the latter case one must take care, as the presence of strong spin orbit coupling can further split the crystal field levels, which in some cases can lead to ambiguous results. To illustrate these points, we take the $4d$ ion Rh$^{4+}$ as an example. In the absence of strong spin-orbit coupling and in an octahedral crystal field, Rh$^{4+}$ with a $4d^5$ electronic configuration would have a single unpaired electron in its $t_{\text{2g}}$ orbitals, giving a local moment with $S = \sfrac{1}{2}$. In the presence of strong spin orbit coupling, such as in Sr$_4$RhO$_6$, the $t_{\text{2g}}$ orbitals are split into an effective $j = \sfrac{3}{2}$ quadruplet and an effective $j = \sfrac{1}{2}$ doublet (Rau \emph{et al.} provide a full description of this effect~\cite{rau2016spin}). In this case, the observed moment may be only slightly enhanced from the $S = \sfrac{1}{2}$ case~\cite{vente1995sr4}. While Curie-Weiss behavior is observed in both of these scenarios, distinguishing between them requires direct spectroscopic evidence of the splitting of the $t_{\text{2g}}$ orbitals~\cite{calder2015spin}. In still other Rh$^{4+}$ materials, such as Sr$_2$RhO$_4$, a metallic state is observed involving the rhodium $4d$ band. Therefore, no local moment is observed and the susceptibility shows only Pauli paramagnetism~\cite{perry2006sr2rho4}. Martins \emph{et al.} provide an excellent survey of how these spin-orbit effects play out in other $4d$ and $5d$ oxides~\cite{martins2017coulomb}.

\section*{Magnetic Frustration}
\label{frustration}

The phrase ``magnetic frustration'' immediately evokes images of unhappy spins, unable to accomplish their goals. This imagery turns out to be essentially correct. Magnetic frustration refers to materials that experience competing interactions that cannot be simultaneously fulfilled, resulting in a large degeneracy of ground state spin configurations. This frustration can arise due to constraints of the lattice geometry (so-called ``geometric frustation'', as observed for the kagome lattice), or competing interactions (as observed in the $J_1$-$J_2$ square lattice), or both. The end result of frustration is a suppression of the magnetic ordering transition as the system struggles to find a suitable compromise and, frequently, a more exotic form of magnetic order or perhaps no discernible order at all. Magnetic frustration is a fascinating topic on which many excellent reviews exist~\cite{ramirez1994strongly,greedan2010geometrically,balents2010spin,lacroix2011introduction,diep2013frustrated,broholm2020quantum}. Here, we will limit our remarks to the key role magnetic susceptibility measurements can play in identifying a magnetically frustrated material.

In order to benchmark whether a material is magnetically frustrated one can apply the Curie-Weiss law to determine its frustration index, $f$, which is defined as:
\begin{equation}
f = \frac{\theta_{\text{CW}}}{T_N} \textrm{~~~~~or~~~~~} f = \frac{\theta_{\text{CW}}}{T_f},
\end{equation}
where $T_{\text{N}}$ and $T_{\text{f}}$ are the temperatures at which the system magnetically orders or freezes, respectively. In cases where no ordering or freezing is observed, the lowest measured temperature (e.g. 1.8~K in a typical $^4$He experiment) can be substituted for $T_{\text{N}}$ to provide a lower bound on the frustration index.
The general idea underlying the frustration index is that the Curie-Weiss temperature $\theta_{\text{CW}}$ parameterizes the strength of the magnetic interactions and, therefore, an unfrustrated material would be expected to order at roughly this temperature, yielding $f = 1$. In reality, typical unfrustrated materials can have $f = 2 - 5$, due to the fact that further neighbour interactions are not incorporated into the derivation of the Curie-Weiss law. Magnets with ordering temperatures that are significantly suppressed by the effects of frustration are typically observed to have $f > 5$ with some materials exceeding $f = 100$~\cite{ramirez1994strongly,greedan2010geometrically}. This rule-of-thumb works best in $3d$ transition metal magnets where $\theta_{\text{CW}}$ values can typically be trusted and higher ordering temperatures are typically expected. Conversely, one must exercise great caution in the same treatment of $4f$ magnets due to crystal field effects that can artificially inflate $\theta_{\text{CW}}$ and the typically lower ordering temperatures.

\section*{Outlook}

Magnetic susceptibility is akin to fingerprinting in the forensic sciences in its ability to reveal the magnetic identity of a material. Over the last century, the increasing ease, availability, and quality of magnetic susceptibility data have made this technique the preferred first step in the characterization of most new magnetic materials. As we have detailed in this tutorial, the information that can be gleaned through this one simple measurement is immense. If the data are treated appropriately, one can determine the magnetic identity of the material, whether it be paramagnetic or diamagnetic, the presence or absence of a magnetic ordering or freezing transitions, anisotropy in single crystal measurements, low-dimensionality or frustration, among others. Technological advances are allowing measurements to be more routinely performed under ever-more extreme conditions, high magnetic field, helium-3 and dilution refrigerator temperatures, high-pressure conditions, and even the combination of the three~\cite{ishizuka1995precise,tateiwa2011miniature,feng2014compact}. These technical advances will further extend the utility of magnetic susceptibility measurements. There can therefore be no doubt that magnetic susceptibility will remain one of the most useful and versatile tools in the characterization of magnetic materials.

\begin{acknowledgments}

This work was supported by the the Natural Sciences and Engineering Research Council of Canada and the CIFAR Azrieli Global Scholars program. This research was undertaken thanks in part to funding from the Canada First Research Excellence Fund, Quantum Materials and Future Technologies Program. 

\end{acknowledgments} 

\bibliography{bibfile}
\end{document}